\documentclass[prd,aps,floats,twocolumn]{revtex4}
\usepackage{epsfig}

\newcommand{\be}{\begin{equation}}
\newcommand{\ee}{\end{equation}}
\newcommand{\ba}{\begin{eqnarray}}
\newcommand{\ea}{\end{eqnarray}}
\newcommand{\bc}{}

\newcommand{\SH}{{\cal{H}}}
\begin{document}

\preprint{\
\begin{tabular}{rr}
&
\end{tabular}
}
\title{On the growth of structure in theories with a dynamical preferred frame}
\author{T.G~Zlosnik$^{1}$, P.G.~Ferreira$^{1}$, 
G.D.~Starkman$^{2}$ }
%
\affiliation{
$^1$Astrophysics, University of Oxford, Denys Wilkinson Building, Keble Road, Oxford OX1 3RH, UK\\
$^{2}$ Department of Physics, Case-Western Reserve University, Cleveland, Ohio, U.S.A.
}

\begin{abstract}
We study the cosmological stability of a class of theories with a dynamical preferred frame. 
For a range of actions, we find cosmological solutions which
are compatible with observations of the recent history of the Universe:
a matter dominated era followed by accelerated expansion. We then 
study the evolution of linear perturbations on these backgrounds and
find conditions on the parameters of the theory which allow for the growth of structure
sourced by the new degrees of freedom.
\end{abstract}

\date{\today}
\pacs{PACS Numbers : }
\maketitle

\section{Introduction}

Theories of modified gravity can be invoked as viable solutions to
the problem of missing mass. Though initially restricted to making predictions
regarding quasistatic, weak field configurations \cite{milgrom,angus,smgaugh}, more recent models
are derivable from generally covariant actions \cite{bek1,navvan,bruneton,dub} 
and so can, in principle, be compared with traditional tests of general relativity
(for instance see \cite{kahya,seif,zhang}).

The best known proposal,
the Tensor-Vector-Scalar (TeVeS) theory of gravity proposed by Bekenstein
has been studied in some depth \cite{bek2,fam,skor,mavro}, and has been found to
explain a plethora of observations, from galactic rotation
curves to the growth of structure during recombination without invoking dark matter. We have shown \cite{zfs1}
that it can be seen
as a particular case of Einstein-Aether theories of gravity \cite{ae} with
non-canonical kinetic terms and that one can consider a different
subclass of these models which seem to be observationally viable \cite{zfs2}.
For convenience and to distinguish them from standard
Einstein-Aether theories, we have named these 
theories Generalized Einstein Aether Theories.

In this paper we wish to explore, in as general terms as possible,
the cosmological stability of these theories. It has been shown 
that the ``vector'' part of TeVeS plays an essential role in the
growth of structure that leads to the formation of galaxies \cite{dodel}. Perturbations
in the time-like vector field are gravitationally unstable and
grow in a manner akin, but not identical, to that of normal, pressureless
matter. We wish to find if this phenomenon is present in Generalized Einstein Aether Theories. 

In this paper we present the following results:
\begin{itemize}
\item we present the full set of background, and linear perturbation,
equations for Generalized Einstein Aether Theories;
\item  for a broad class of models there is accelerated
expansion following a period of matter domination;
\item  in a pure baryonic universe, the growth of structure
can be sustained and enhanced by the presence of the Aether vector field;
\item although the growth of structure is driven by the
vector field, unlike in TeVeS, it remains strictly unit timelike leading to
qualitatively different features;
\item there are choices of parameters in the Einstein-Aether theories that
give a reasonable fit to current measurements of large scale structure
\item a few notable features can be used to distinguish it from conventional
gravity and dark matter, in particular a different growth rate for structure and
the presence of non zero, scalar anisotropic stresses (``$\Phi-\Psi$'').
\end{itemize}

There are a number of steps we must go through before we can assess
if there is the desired instability of the vector field, or aether, in these
theories. We are looking for it in the cosmological setting and hence
we must identify the restricted class of theories that give us
acceptable cosmologies. We do so in Section \ref{sect:background}. Once we
have established what region of parameter space we want to explore,
we then check if perturbations in the vector field grow in a matter dominated
era. We shall see, in Section
\ref{sect:vector} that this defines a subset of parameters.
We then consider the full coupled system of metric, matter, and vector field
perturbations for modes outside the horizon well before recombination and
trace their evolution up to the present time.

As will become clear, we are able to identify the subset of theories
which could possibly lead to realistic growth of structure in a purely baryonic
universe. In Section \ref{sect:conc} we discuss if the constraints we find
 in this paper are compatible with a host of constraints
that have been identified in \cite{zfs2}.
\section{The theory}
Let us first recap the structure of Generalized Einstein Aether Theories.
A general action for a vector field, $A^{\mu}$ coupled to gravity can
be written in
the form
\begin{eqnarray}
S=\int d^4x \sqrt{-g}\left[\frac{1}{2}M_{PL}^{2}R+{\cal L}(g^{\mu\nu},A^{\nu})\right]
+S_{M}
\end{eqnarray}
where $g_{\mu\nu}$ is the metric, $R$ is the Ricci scalar of that metric,
$M_{PL}=\frac{1}{\sqrt{8\pi G}}$ is the Planck mass,
$S_M$ is the matter action, and $\cal{L}$ is constructed to be generally
covariant and local. $S_M$ couples only to the metric, $g_{\mu\nu}$ and {\it not}
to $A^{\mu}$. Throughout this paper, repeated Greek indices are understood to be summed over
0..3 and Latin indices are summed over 1..3.
{\cal F}urthermore we will consider for
most of this paper that $A^{\mu}$ is unit time-like.

In this paper we will focus on the subset of
theories proposed in \cite{zfs2}, i.e. we consider a Lagrangian that
only depends on covariant derivatives of $A^{\mu}$ and
the time-like constraint. It can be written
in the form
\be
\label{eq:Lagrangian}
{\cal L}(g^{ab},A^{a})=\frac{1}{2}M^2M_{PL}^{2}
 	 {\cal F}({\cal K}) +\frac{1}{2}M_{PL}^{2}\lambda(A^\alpha A_\alpha+1)  ,
\ee
where
\begin{eqnarray}
{\cal K}&\equiv&M^{-2}{\cal K}^{\alpha\beta}_{\phantom{\alpha\beta}\gamma\sigma}
\nabla_\alpha A^{\gamma}\nabla_\beta A^{\sigma} ,\nonumber\\
{\cal K}^{\alpha\beta}_{\phantom{\alpha\beta}\gamma\delta}&\equiv&c_1g^{\alpha\beta}g_{\gamma\sigma}
+c_2\delta^{\alpha}_{\phantom{\alpha}\gamma}\delta^{\beta}_{\phantom{\beta}\sigma}+
c_3\delta^{\alpha}_{\phantom{\alpha}\sigma}\delta^{\beta}_{\phantom{\beta}\gamma} .\nonumber
\end{eqnarray}
Here $c_i$ are dimensionless constants and $M$
has the dimension of mass. $\lambda$ is a non-dynamical Lagrange-multiplier
field with dimensions of mass-squared. One may additionally consider a `$c_{4}$' term in $K_{\alpha\beta}^{\phantom{\alpha\beta}\mu\nu}$ proportional to 
$g_{\alpha\beta}A^{\mu}A^{\nu}$ \cite{ae}.
In \cite{zfs2} we considered the limit where matter sources in the gravitational field equations are negligible. We found that for the perturbed
aether degrees of freedom to have a positive Hamiltonian, and to avoid superluminal propagation in spin-0 aether perturbations and gravitational waves collectively implied that: 

\begin{equation}
\label{ciconditions}
c_1<0, \quad c_2\leq0, \quad  {\rm and} \quad c_1+c_2+c_3\leq0
\end{equation}

The gravitational field equations are
\begin{equation}
G_{\alpha\beta}=\tilde{T}_{\alpha\beta}+8\pi GT^{matter}_{\alpha\beta}
\label{fieldI}
\end{equation}
where the stress-energy tensor for the vector field is given by
\begin{eqnarray}
\tilde{T}_{\alpha\beta} &=& \frac{1}{2}\nabla_{\sigma}
({\cal F}_{{\cal K}}(J_{(\alpha}^{\phantom{\alpha}\sigma}A_{\beta)}-
J^{\sigma}_{\phantom{\sigma}(\alpha}A_{\beta)}-J_{(\alpha\beta)}A^{\sigma}))
\nonumber \\ && -{\cal F}_{{\cal K}}Y_{(\alpha\beta)}
+\frac{1}{2}g_{\alpha\beta}{\cal F}+\lambda A_{\alpha}A_{\beta} ,
\end{eqnarray}
where
\begin{eqnarray}
{\cal F}_{{\cal K}} &\equiv& \frac{d{\cal F}}{d{\cal K}} , \\
J^{\alpha}_{\phantom{\alpha}\sigma} &\equiv&
(\cal{K}^{\alpha\beta}_{\phantom{\alpha\beta}\sigma\gamma}+
\cal{K}^{\beta\alpha}_{\phantom{\beta\alpha}\gamma\sigma})\nabla_{\beta}A^{\gamma} 
\end{eqnarray}
and
\be
Y_{\alpha\beta} = c_{1}[\nabla^{\xi}A_{\alpha}\nabla_{\xi}A_{\beta}-\nabla_{\alpha}A^{\xi}\nabla_{\beta}A_{\xi}] .
\ee
Brackets around indices denote
symmetrization.

The equations of motion for the vector field are
\begin{eqnarray}
\nabla_{\alpha}({\cal F}_{{\cal K}}J^{\alpha}_{\phantom{\alpha}\beta})
&=&2\lambda A_{\beta}
\label{vectoreom}
\end{eqnarray}
Variations of $\lambda$ will fix $A^\mu A_\mu=-1$. Transvecting the equation
of motion for $A_{\mu}$ with $A^{\mu}$ allows us to solve for the Lagrange
multiplier:
\begin{eqnarray}
\lambda=-\frac{1}{2}A^{\beta}\nabla_{\alpha}({\cal F}_{{\cal K}}J^{\alpha}_{\phantom{\alpha}\beta}) 
\end{eqnarray}

\section{Background Evolution} \label{sect:background}

We first consider the case of a  homogeneous and
isotropic universe in which the metric is of the form
\be
g_{\mu\nu}dx^{\mu}dx^{\nu} =-{dt}^{2}+a(t)^{2}\delta_{ij}dx^{i}dx^{j} ,
\ee

where $t$ is physical time and $a(t)$ is the scale factor. 
These solutions will be used as the background
on which we can study the growth of perturbations.
The vector field must respect the spacial homogeneity and isotropy of the system and so will only have a non-vanishing `$t$' component; the constraint
fixes $A^{\mu}= (1,0,0,0)$. The energy-momentum tensor of the
matter is of the form 
\be 
{T}^{matter}_{\alpha\beta} = \rho U_{\alpha}U_{\beta}
  +P(g_{\alpha\beta} + U_\alpha U_\beta)  ,
\ee
where $\rho$ is the matter energy density, $P$ is pressure 
and we have introduce a unit timelike four-vector $U^{\nu}$,
{\it i.e.} satisfying $g_{\mu\nu}U^{\mu}U^{\mu} = -1$.

The equations of motion simplify dramatically with these symmetries and we
find that:
\begin{eqnarray}
\label{KappaofH}
\nabla_{\mu}A^{\mu}&=& 3H\\
{\cal K} &=& 3\frac{\alpha H^{2}}{M^{2}} 
\end{eqnarray}
where $H \equiv \frac{\dot{a}}{a}$,
the dot denotes differentiation with respect to $t$, and,
following \cite{CL}, we define $\alpha = c_{1}+3c_{2}+c_{3}$. Recall that
$\alpha$ is negative and hence so is ${\cal K}$.

The modified Einstein's equations now become:
\begin{eqnarray}
[1-{\cal F}_{{\cal K}}\alpha]H^{2}+\frac{1}{6}{\cal F}M^{2}& = &\frac{8\pi G}{3}\rho
\\
-[1-2\alpha{\cal F}_{{\cal K}}]H^2-2[1-\frac{1}{2}\alpha{\cal F}']\frac{\ddot a}{a}+\nonumber\\
{\dot {\cal F}}_{{\cal K}}\alpha H-\frac{1}{2}{\cal F}M^2 &=& 8\pi G P  .
\end{eqnarray}
We can rewrite these equations in a different form:
\begin{eqnarray}
\left[1-\alpha{\cal K}^{1/2}\frac{d}{d{\cal K}}\left(\frac{\cal F}{{\cal K}^{1/2}}\right)\right]H^2 &=& \frac{8\pi G}{3}\rho \label{00m} \\
\frac{d}{dt}(-2H+{\cal F}_{{\cal K}}\alpha H)&=&8 \pi G (\rho+P) .\label{summ}
\end{eqnarray}

As discussed in \cite{zfs2}, the form of ${\cal F}$ for negative values
of ${\cal K}$ cannot be constrained by looking at the structure of the
modified force law near spherically symmetric bodies or near the
Newtonian regime where ${\cal K}>0$. Hence we can consider any general form. Initially it
is informative to restrict ourselves to 
\begin{eqnarray}
\label{eqn:FofK}
{\cal F}=\gamma (-{\cal K})^n .
\end{eqnarray}
With such
a choice, we find that equation \ref{00m} is more clearly a modified  Friedmann equation:
\begin{eqnarray}
\label{eq:modfri}
\left[1+\epsilon\left(\frac{H}{M}\right)^{2(n-1)}\right]H^2=\frac{8\pi G}{3}\rho  ,
\end{eqnarray}
where 
\be
\epsilon=(1-2n)\gamma(-3\alpha)^{n}/6. 
\ee

For the purpose of this paper and as can be seen from the Friedmann equation,
we have restricted ourselves to flat universes. 
This implies that there is a relationship between
\be 
\Omega_{m}\equiv8\pi g\rho_{0}/3H_0^2
\ee
(where $H_0$ is the Hubble constant today) and $\gamma$. Indeed, 
\begin{eqnarray}
\gamma=\frac{6(\Omega_{m}-1)}{(1-2n)(-3\alpha)^n}\left(\frac{M}{H_0}\right)^{2(n-1)} .
\end{eqnarray}

We can immediately identify a few special cases:
\begin{itemize}
\item $n=1/2$ -- the Friedmann equations are unchanged ($\epsilon=0$) and there is no
effect on the background cosmology; 
\item $n=1$ -- we have that 
$\epsilon=\gamma\alpha/2$ and recover the
results of \cite{CL} ($\gamma\leq 0$), i.e. Newton's constant is rescaled,
$G'=G/(1+\epsilon)$; 
\item $n=0$ --  we recover a cosmological constant,
$\Lambda\simeq sign(-\gamma)  M^2$.
\end{itemize}

More generally we will obtain different regimes depending on the relative size
of each term in equation \ref{eq:modfri}. Consider $\epsilon<0$. We can define
$\epsilon=-{\tilde \epsilon}^2$ to rewrite the modified Friedmann equation
as
\begin{eqnarray}
\label{eq:modfrib}
\left[H-{\tilde \epsilon}\left(\frac{H}{M}\right)^n M\right]
\left[H+{\tilde \epsilon}\left(\frac{H}{M}\right)^n M\right]=\frac{8\pi G}{3}\rho .
\end{eqnarray}
Clearly $H\rightarrow H_{eq}$ where $H_{eq}=M{\tilde \epsilon}^{1/(1-n)}$. We
can expand around the equilibrium solution, $H\simeq H_{eq}(1+h)$ to find the
rate of approach to equilibrium given by
\begin{eqnarray}
h=\frac{1}{2(1-n)}\frac{8\pi G \rho}{3H_{eq}^2} .
\end{eqnarray}
In this regime 
accelerated expansion is approached as $\rho\rightarrow 0$. Note that
we have assumed that $\epsilon<0$ so that we have $n<\frac{1}{2}$ and $\gamma<0$
or $n>\frac{1}{2}$ and $\gamma>0$. Note also that this corresponds
to $\Omega<1$.

There is an important qualitative difference in the evolution of 
$H$ depending on the value of $n$.
If $n<1$, then the exponent of $H/M$ in equation  (\ref{eq:modfri})
(and in equation (\ref{eq:modfrib})) is smaller than the exponent of the standard
$H$ term. The $H/M$ terms will therefore contribute most  when $\vert H/M\vert$ is small. 
With $H$ currently positive and smaller than in the past, this means that we could
be approaching a phase of constant $H$.  On the other hand, if $n>1$, then the
$H/M$ terms will contribute most  when $\vert H/M\vert$ is large, in other words
in the past.  Since the expansion is observed to be approaching a period of approximately constant $H$
($w\simeq-1$), if $\epsilon<0$ we prefer $n<1$.

We can now consider $\epsilon>0$. For this regime there is again a
competition between the conventional $H^2$ term and the modification
due to the vector field. It is still the case that if  $n>1$ any modifications
in the background cosmology come into play
for high values of $H/M$, i.e. at early times. As claimed in \cite{zfs2},
this can be a source of cosmological inflation in the early universe.
This is not the regime we are interested in, in this paper. 

For $0<n<1$ we find that the expansion rate at late times will be retarded. If we assume
$\rho\propto a^{-3}$ we find that $a\propto t^{2n/3}$ with the
onset of the regime occurring when $H\simeq \epsilon^{-1/2(n-1)}M$.
Note that as $\epsilon \to 0$, this will occur as $t\rightarrow \infty$.
Holding $\gamma$ and $\alpha$ fixed, this means that as $n\to 1/2$, $t\rightarrow \infty$.
Also, since $H \simeq 2n/3$ in this regime, as $n$ tends to $0$ the expansion rate decreases. 

For $n=0$ we find that the equations correspond to an Anti-deSitter spacetime
with additional matter; the Universe will expand, slow down and then loiter
at $H\simeq0$ and constant $a$. This point triggers the onset of an exotic
range of behaviour, for $n<0$. For negative values of $n$ we find that
$a$ will expand and $H$ will tend to $\epsilon^{1/2n}M$ after which $H$ changes sign discontinuously and $a$ contracts. This is signalled by ${\dot H}\rightarrow \infty$ at the transition. The energy momentum tensor transits smoothly
at this point, guaranteeing that the system is well behaved.

\begin{figure}
\epsfig{file=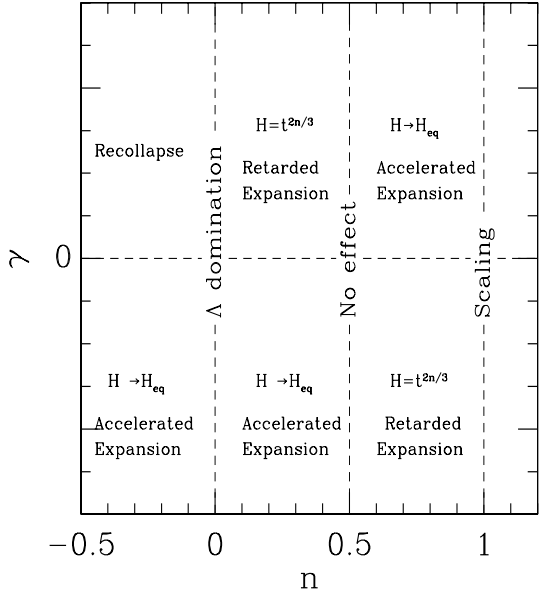,width=8.3cm,height=9.3cm}
\caption{A schematic representation of the types of the late-time background evolution as a function of $(n,\gamma)$ for $n<1$.}
\label{fig1}
\end{figure}

We can summarize the behaviour in a plot in $(n,\gamma)$ in Figure \ref{fig1}. 
The region of interest to us is clearly $\gamma<0$ and $n<1/2$ and $\gamma>0$ and $n>1/2$. In this region we can envisage, for example, a period of matter domination after recombination followed by late time accelerated expansion.

\section{Linear perturbation Theory}

We now look at linear perturbations on the homogeneous
background. Throughout this paper we will focus on scalar
perturbations and work in the
conformal Newtonian gauge. The metric can be expanded
as
\begin{equation}
g_{\mu\nu}dx^{\mu}dx^{\nu}=-(1+2\epsilon\Psi)dt^{2}+a^{2}(1-2\epsilon\Phi)\delta_{ij}dx^{i}dx^{j} ,
\end{equation}
while the four vector $A^{a}$ can be expanded as
\begin{eqnarray}
A^{\mu}&=&(1+\epsilon X,\epsilon\partial^{j}Z)\\
\nonumber &=& (1+\epsilon X,\frac{\epsilon}{a^{2}}\partial_{j}Z) .
\end{eqnarray}
Here indices have been lowered with the background metric. In Fourier space:
\begin{equation}
A^{\mu}=(1-\epsilon \Psi,i\frac{\epsilon}{a} k_{j}V),
\end{equation}
where, for computational convenience, we have defined $V\equiv Z / a$ and have
used the fact that the constraint fixes $X=-\Psi$.

We can expand ${\cal K}$ to linear order:
\begin{eqnarray}
{\cal K} &=& {\cal K}^{0}+\epsilon {\cal K}^{\epsilon} \\
{\cal K}^{0} &=& 3\frac{\alpha H^{2}}{M^{2}}\\
{\cal K}^{\epsilon} &=& -2\frac{\alpha H}{M^{2}}(k^{2}\frac{V}{a}+3H\Psi+3\dot{\Phi}) 
\end{eqnarray}

We now switch to the conformal Newtonian gauge. 
Derivatives with respect to conformal time, $\eta$, are denoted by $'$ and $\frac{a'}{a}\equiv {\cal H}$. 
The evolution equation for the perturbation in the vector field becomes
\begin{eqnarray}
\label{vecp}
0 &=& c_{1}[V''+k^{2}V+2{\cal H}V'+2{\cal H}^{2}V+\\ 
\nonumber && +\Psi'+\Phi '+2{\cal H}\Psi] \\
\nonumber && +c2[k^{2}V+6{\cal H}^{2}V-3\frac{a''}{a}V+3\Phi'+3{\cal H}\Psi]\\
\nonumber && +c3[k^{2}V+2{\cal H}^{2}V-\frac{a''}{a}V+\Phi'+{\cal H}\Psi]\\
\nonumber && +\frac{F_{KK}}{F_{K}}[-K^{\epsilon}\alpha\SH\\
\nonumber &&-K^{0'}(-c_{1}(V'+\Psi)+3c_{2}\SH V+c_{3}\SH V)] .
\end{eqnarray}

The perturbation in the vector field is sourced by the two gravitational
potentials $\Phi$ and $\Psi$. These in turn are
seeded by the vector field and the perturbations in the matter field.
The first order perturbations to the vector field's stress energy tensor are:
%
%
\begin{eqnarray}
a^{2}\delta \tilde{T}^{0}_{\phantom{0}0}&=& F_{K}c_{1}[-{\cal H}k^{2}V-k^{2}V'-k^{2}\Psi]  \\
\nonumber && + F_{K}\alpha[2{\cal H}k^{2}V+6{\cal H}\Phi '+6{\cal H}^{2}\Psi] \\
\nonumber && +\frac{1}{2}a^{2}F_{K}M^{2} K^{\epsilon}-3 F_{KK}\alpha{\cal H}^{2}K^{\epsilon}  \\
\nonumber &=& F_{K}c_{1}[-{\cal H}k^{2}V-k^{2}V'-k^{2}\Psi]  \\
\nonumber && +F_{K}\alpha(2n-1)[{\cal H}k^{2}V+3{\cal H}\Phi '+3{\cal H}^{2}\Psi] ,
\end{eqnarray}
%
%
\begin{eqnarray}
a^{2}\delta \tilde{T}^{0}_{\phantom{0}i} &=& ik_{i}F_{K}c_{1}[V''+2{\cal H}V'+\frac{a''}{a}V  \\ 
\nonumber && +\Psi '+{\cal H}\Psi] \\
\nonumber && +ik_{i}F_{K}\alpha [2{\cal H}^{2}V-\frac{a''}{a}V]\\
\nonumber && +ik_{i}F_{KK}{\cal K}^{0'}[c_{1}({\cal H}V+V'+\Psi) \\
\nonumber && - \alpha {\cal H}V] ,
\end{eqnarray}
%
%
\begin{eqnarray}
\label{eq:tij}
a^{2}\delta \tilde{T}^{i}_{\phantom{i}j} &=& F_{K}c_{2}k^{2}[2{\cal H}V+V']\delta^{i}_{\phantom{i}j} \\
\nonumber && F_{K}\alpha[k^{2}{\cal H}V+5{\cal H}\Phi'+\Phi ''\\
\nonumber && + 2\frac{a''}{a}\Psi+2{\cal H}^{2}\Psi+{\cal H}\Psi']\delta^{i}_{\phantom{i}j}\\
\nonumber &&+F_{K}(c_{1}+c_{3})[2{\cal H}V+V']k_{i}k_{j} \\
\nonumber &&+\frac{1}{2}a^{2}F_{K}M^{2}K^{\epsilon}\delta^{i}_{\phantom{i}j}\\
\nonumber &&+F_{KK}[-\alpha K^{\epsilon}\frac{a''}{a}-(c_{1}+c_{2}+c_{3}) K^{\epsilon}{\cal H}^{2}\\
\nonumber &&-\alpha {\cal H}K^{\epsilon '}+\alpha K^{0 '}\Phi '+2\alpha K^{0 '}{\cal H}\Psi\\
\nonumber && -\alpha \ln(F_{KK})'K^{\epsilon}{\cal H}+c_{2}K^{0'}k^{2}V]\delta^{i}_{\phantom{i}j}\\
\nonumber && +(c_{1}+c_{3})F_{KK}K^{0'}Vk_{i}k_{j}  .
\end{eqnarray}
where the second expression for $a^{2}\delta\tilde{T}^{0}_{\phantom{0}0}$ assumes the monomial form for $F({\cal K})$.
%
In the absence of anisotropic stresses in the matter fields, we may obtain an algebraic relation between the metric potentials $\Phi$ and $\Psi$ by computing the transverse, traceless part of the perturbed Einstein equations. i.e.:
\begin{eqnarray}
k^{2}(\Psi-\Phi) &=& \frac{3}{2}a^{2}(\hat{k}_{i}\hat{k}_{j}-\frac{1}{3}\delta_{ij})(\delta\tilde{T}^{i}_{j})  \label{GRTT} \\
\nonumber &=& (c_{1}+c_{3})k^{2}[F_{K}(2{\cal H}V+V')+F_{KK}K^{0'}V] .
\end{eqnarray}
We can see then that generally $\tilde{T}^{i}_{\phantom{i}j}=A \delta^{i}_{\phantom{i}j}$  if $c_{1}+c_{3}=0$. 
From the perturbed Einstein tensor we obtain:
\begin{eqnarray}
a^{2}[G^{0}_{\phantom{0}0}-\frac{3}{k^{2}}{\cal H}ik_{j}G^{0}_{\phantom{0}j}]=2k^{2}\Phi .
\end{eqnarray}
Similarly,
\begin{eqnarray}
a^{2}[\tilde{T}^{0}_{\phantom{0}0}-\frac{3}{k^{2}}{\cal H}ik_{j}\tilde{T}^{0}_{\phantom{0}j}]=
-F_{K}c_{1}k^{2}[V'+\Psi+(3+2\tilde{c}_{3}){\cal H}V] ,
\end{eqnarray}
where we have used the identity provided by (\ref{vecp}). 

We take the dominant remaining contributions to be from the baryons and radiation, treated as perfect fluids. No contribution from anisotropic shear is considered. Therefore, the perturbations to a component described by stress energy tensor  \textbf{T} are given by:
\begin{eqnarray}
\label{deltaT}
\delta T^{0}_{\phantom{0}0(a)} &=& -\bar{\rho}_{(a)}\delta_{(a)} \\
\delta T^{0}_{\phantom{0}i(a)} &=& [\bar{\rho}_{(a)}+\bar{P}_{(a)}]v_{i(a)} = -\delta T^{i}_{\phantom{i}0} \\
\delta T^{i}_{\phantom{i}j(a)} &=& [\bar{P}_{(a)}+\delta P_{(a)}]\delta^{i}_{\phantom{i}j} 
\end{eqnarray}
where $\bar{\rho}_{(a)}$ and $\bar{P}_{(a)}$ are  the background energy density and pressure of the fluid, $\delta_{(a)}$ is the density contrast $\frac{\delta\rho_{(a)}}{\bar{\rho}_{(a)}}$, $\delta P_{(a)}$ is the perturbation to the fluid pressure,  and $v^{i}_{(a)}$ are components of the fluid's 3 velocity. We also define the divergence of the fluid velocity, $\theta=i k_{i}v^{i}$. Calculating the contribution from this tensor in Einstein's equations, we find the following expression for Poisson's equation:
\begin{eqnarray}
\label{eqn:Poisson}
k^{2}\Phi&=&-\frac{1}{2}F_{K}c_{1}k^{2}[V'+\Psi+(3+2\tilde{c}_{3}){\cal H}V]\\
\nonumber && -4\pi Ga^{2}\sum_{a}(\bar{\rho}_{a}\delta_{a}+3(\bar{\rho}_{a}+\bar{P}_{a}){\cal H}\frac{\theta_{a}}{k^{2}}) .
\end{eqnarray}

\section{Evolution of Vector Perturbations} \label{sect:vector}

We now investigate solutions to the vector perturbation equation (\ref{vecp}).
We first establish the conditions for the existence of growing modes during the matter dominated epoch, 
since in their absence the low amplitude of perturbations in the matter at last scattering, 
and the subsequent effects of Silk damping, will result in a failure to evolve structure in the universe.

We consider therefore that we are following the evolution of the vector field modes during an epoch
when the scale factor $a(\eta) \propto \eta^m$. $m=2$ during the matter-dominated
epoch.  We also suppose that  $F({\cal K})$ is dominated by a single power law -- as above
$F({\cal K}) = \gamma \left(-{\cal K}\right)^n$ -- and that this form is consistent with a period of radiation domination
followed by matter domination. 
In this case the source-free part of equation (\ref{vecp}) is:
\begin{eqnarray}
\label{eqn:sourcefree}
0 &=&  V''+b_{1}\frac{1}{\eta}V'+(b_{2}+b_{3}(k\eta)^{2})\frac{1}{\eta^{2}}V ,
\end{eqnarray}
where 
\begin{eqnarray}
b_{1} &\equiv& 2m-(2m+2)(n-1) \\
b_{2} &\equiv& (m^{2}+m+2m(n-1)(m+1))\tilde{\alpha}\\ 
\nonumber && +(m^{2}-m-2m(n-1)(m+1)) \\
b_{3} &\equiv& 1+\tilde{c}_{2}+\tilde{c}_{3}+\frac{2}{3}(n-1)\tilde{\alpha} \label{bees}.
\end{eqnarray}
Here $\tilde{c}_i\equiv c_i/c_1$, and $\tilde{\alpha}\equiv\alpha/c_1$. 
Because of the condition (\ref{ciconditions}), $\tilde{c_2}\geq0$,
and $1+\tilde{c}_2+\tilde{c}_3\geq0$.  Consequently $\tilde{\alpha}\geq0$ also.

In addition to the source-free terms included in  (\ref{eqn:sourcefree}), 
the vector evolution equation (\ref{vecp}) contains both 
self-gravity terms and terms sourced by matter perturbations.
Since we are interested here in the initial growth of structure, 
early in the matter dominated epoch, we may ignore the matter perturbations.
What then of the self-gravity terms in (\ref{vecp})?  To lowest order (in $\epsilon$) these
may be collected into the following instructive form (and divided by $c_1$) in which they
can be compared to the terms of equation (\ref{eqn:sourcefree}):
\begin{eqnarray}
\label{eqn:vectorselfgravity}
\left(\Psi'-\Phi'\right) + \left(1+\tilde\alpha\left(2n-1\right)\right)\left(\Phi'+
{\cal H}\Psi\right) \\ 
+2\left(n-1\right)\left(2\tilde\alpha - \frac{m+1}{m}\frac{1}{c_1}\right){\cal H}\Psi .
\end{eqnarray}

From (\ref{GRTT}) we see that the first term $\left(\Psi'-\Phi'\right)$
is proportional to ${\cal F}_K\equiv d{\cal F}/dK$ and its derivatives with resepct to  $K$ and $\eta$.
Recall that ${\cal F}=\gamma (-{\cal K})^n$  (equation (\ref{eqn:FofK})), and 
that we have found that interesting  background cosmology suggests that 
(at least at late times) we confine our attention to $n<1$.  
Equation (\ref{KappaofH}),
\begin{equation}
{\cal K} = 3\frac{\alpha H^{2}}{M^{2}} , 
\end{equation}
and the fact that in order to recover galaxy rotation curves  we must have
$M$ be of order the current Hubble parameter $H_0$  (\cite{zfs2}), 
together imply that unless $\vert\alpha\vert\ll1$, ${\cal K}$ will be  large
at large redshift when $H\gg H_0$.
Thus, in the period of interest soon after matter domination,  
we expect $F_K$ and its derivatives to be very small,
and so the first term in (\ref{eqn:vectorselfgravity}) to be negligible.

The second term in (\ref{eqn:vectorselfgravity}) is proportional to $\left(\Phi'+{\cal H}\Psi\right)$.
According to the $G^{0}_{\phantom{0}j}$ Einstein equation 
\be
k^2\left(\Phi'+{\cal H}\Psi\right) = \frac{3}{2}{\cal H}^2(1+3w)\theta + i a^2 k^j \delta \tilde{T}^{0}_{\phantom{0}j} .
\ee
We are specifically interested in the growth of vector field modes at times when matter seeds are small, 
so we can set $\theta=0$.  Meanwhile $\delta \tilde{T}^{0}_{\phantom{0}j}$ is, according to equation  (\ref{deltaT}),
proportional to ${\cal F}_K$ and its derivatives.  As we have just seen, such terms are suppressed
at early times.

Finally, the third term in equation (\ref{eqn:vectorselfgravity}) is proportional to $\Psi$.  
Using the Poisson equation (\ref{eqn:Poisson}), and equation (\ref{GRTT}), one can reexpress $\Psi$:
\begin{eqnarray}
\Psi \left(1 + \frac{1}{2}{\cal F}_K c_1\right) &=& \frac{1}{2}{\cal F}_K\left(c_1+2c_3\right)V' \\
\nonumber &&  +{\cal F}_K {\cal H} [\frac{1}{2}\left(c_1+2c_3\right)\\
\nonumber && - 2 \left(c_1+c_2\right)\frac{m+1}{m}\left(n-1\right)]V \\
\nonumber &&  -\frac{4\pi G a^{2}\bar{\rho}}{k^{2}}[\delta+3\frac{{\cal H}\theta}{k^{2}}] .
\end{eqnarray}

Again, all terms on the right hand side are either suppressed by factors proportional to ${\cal F}_K$,
or are sourced by matter.

We thus conclude that for an analysis of the existence of vector growing modes
soon after the onset of matter domination in the absence of matter seeds,
we can confine ourselves to the source-free equations (\ref{eqn:sourcefree}).

We can solve analytically the source free vector perturbation equation 
(\ref{eqn:sourcefree}) even with the inclusion of the inhomogeneous terms so long as we
confine ourselves to a particular cosmological epoch, i.e to fixed $m$.  
The solution is:
\begin{eqnarray}
\label{eq:vecev}
V(\textbf{k},\eta) &=& \eta^{\frac{1}{2}(1-b_{1})}[f_{1}(\textbf{k})J(\beta,\sqrt{b_{3}}k\eta)\\
\nonumber &&
+f_{2}(\textbf{k})Y(\beta,\sqrt{b_{3}}k\eta)] ,
\end{eqnarray}
where $J$ and $Y$ are Bessel $J$ and $Y$ functions respectively and the $f_{i}$ are functions to be fixed
by boundary conditions.

We look first for the homogeneous ($k\eta=0$) solution to (\ref{eqn:sourcefree}).
If $\beta \geq 0$, the appropriate form of a solution is $V \propto \eta^p$, hence
\begin{eqnarray}
\label{eq:powerlaw}
V(\eta)= C_{1}\eta^{\frac{1}{2}(1-b_{1})-\beta}+C_{2}\eta^{\frac{1}{2}(1-b_{1})+\beta} .
\end{eqnarray}
The $C_{i}$ are constants of integration and for convenience we have defined $\beta \equiv \frac{1}{2}(1-4 b_{2}
-2 b_{1}+b_{1}^{2})^{\frac{1}{2}}$.

A growing mode therefore exists if and only if
\begin{equation}
\label{growingmodecondition}
\frac{1}{2}(1-b_{1})+\beta > 0 .
\end{equation}

In the matter era (m=2), this condition describes three regimes where growing modes occur:
\begin{eqnarray}
0< \tilde{\alpha}  &<& \frac{6n-7}{3(2n-1)}    \quad{\rm for}\quad  7/6<n\\
\tilde{\alpha}  &>&  \frac{6n-7}{3(2n-1)}   \quad{\rm for}\quad  n < 1/2  .\\
\end{eqnarray}
where we have insisted that $\tilde{\alpha}\geq 0$. 

In the radiation era (m=1), alternative regimes lead to growing modes:

\begin{eqnarray}
0< \tilde{\alpha}  &<& \frac{2n-2}{(2n-1)}    \quad{\rm for}\quad  1<n\\
\tilde{\alpha}  &>&  \frac{2n-2}{(2n-1)}   \quad{\rm for}\quad  n < 1/2  .\\
\end{eqnarray}

If $n$ is unchanged from the radiation-dominated era to the
matter dominated era, then the requirement of growing modes
in the matter dominated  automatically results in growing
modes in the radiation-dominated era.

We first consider the scaling value ($n=1$) previously
well-explored by Jacobson, Mattingly and collaborators. 
Indeed in this case we have that simply that:

\begin{eqnarray}
b_{1}&=& 4 \\
b_{2} &=& 6\tilde{\alpha}+2 \\
\beta &=&\left(\frac{1}{4}-6\tilde{\alpha}\right)^{\frac{1}{2}}
\end{eqnarray}
during the matter era and

\begin{eqnarray}
b_{1} &=& 2 \\
b_{2} &=& 2\tilde{\alpha}\\
\beta &=&\left(\frac{1}{4}-2\tilde{\alpha}\right)^{\frac{1}{2}}
\end{eqnarray}
during the radiation era.

We find decaying solutions of the type (\ref{eq:powerlaw})  for $0\leq\tilde{\alpha}<1/24$ during the matter era and $0\leq\tilde{\alpha}<1/8$ during the radiation era and that, signalling the transition $\beta$ to an imaginary value , greater respective values of $\tilde{\alpha}$ lead to decaying oscillatory modes. 

For more general $F({\cal K})$, only in the regime  ($n<1/2$) do we both find a growing mode, and an approach to 
a late time acceleration in the background. Therefore we will restrict our attention to these values of $n$ for the remainder of the paper.

\section{Evolution of coupled perturbations during the matter era}

We now consider the effect of the vector field during matter domination;
in doing so we take the dominant remaining contribution to the energy density to be baryonic, treated as a pressureless perfect fluid with energy-momentum tensor \textbf{T}.

$\nabla_{a}\delta T^{ab}=0$ produces the following:
\begin{eqnarray}
\delta ' &=& -\theta + 3\Phi '  \label{barev1}\\
\theta ' &=& -{\cal H}\theta + k^{2}\Psi \label{barev2}.
\end{eqnarray}
Furthermore we introduce the variable:
\begin{eqnarray}
V' \equiv E 
\end{eqnarray}

Examining (\ref{vecp}) we expect a first order evolution equation for E. Therefore, the perturbation variables are: $\delta,\theta,\Psi,\Phi,V,E$. In addition we have two constraints between the variables, given by:
\begin{eqnarray}
k^{2}\Phi&=&-\frac{1}{2}F_{K}c_{1}k^{2}[E+\Psi+(3+2\tilde{c}_{3}){\cal H}V]\\
\nonumber && -4\pi Ga^{2}\bar{\rho}[\delta+3{\cal H}\frac{\theta}{k^{2}}]\\
k^{2}(\Psi-\Phi) &=& (c_{1}+c_{3})k^{2}[F_{K}(2{\cal H}V+E)\\
\nonumber && +F_{KK}K^{0'}V].
\end{eqnarray}

Following \cite{bma} we choose to eliminate $\Psi$ via the constraint equations and
integrate the remaining variables. 
We use the $G^{0}_{\phantom{0}0}$ component of Einstein's equations as a first order evolution
equation for $\Phi$ and reserve Poisson's equation as a consistency check.

We can now examine the evolution of perturbations in the matter era after recombination.
In solving the full system of equations numerically, we have found that the solution (\ref{eq:powerlaw})  describes the evolution of $V$ during the matter era for universes that produce a realistic description of the matter power spectrum today, vindicating our choice to neglect the effective metric and matter sources to the vector field equation during the matter era. 

For ease of illustration we will initially consider only the case where $V$ is described by a growing monomial, {\it i.e.}
\begin{equation}
\label{eq:simple}
V=V_{0}\left(\frac{\eta}{\eta_{0}}\right)^{p} ,
\end{equation}

Taking the vector field to be described by (\ref{eq:simple}), 
the arguments from Section \ref{sect:vector} may be similarly used to show that, concomitantly, 
the components of vector field stress energy tensor $\delta\tilde{T}^{a}_{\phantom{a}b}$  
will be dominated by those proportional to $V$ and $E$. 
It follows then that during the matter era both the quantity $a^{2}\delta\tilde{T}^{0}_{\phantom{0}0}$ 
and the difference between the conformal Newtonian gauge potentials take particularly simple forms:
\begin{eqnarray}
a^{2}\delta T^{0}_{\phantom{0}0} &\simeq& -l_{E}\xi(k)k^{2}\eta^{5+p-6n} \label{too} \\
k^{2}(\Psi-\Phi) &\simeq& -l_{S}\xi(k)k^{2}\eta^{5+p-6n} , \label{sifif}\\
\end{eqnarray}
where 
\begin{eqnarray}
\xi(k) &\sim& \gamma V_{0}(k)\left(\frac{1}{\eta_{0}}\right)^{p}k_{hub}^{6-6n}\left(3\alpha\Omega_{m}\left(\frac{H_{0}}{M}\right)^{2}\right)^{n-1} \label{xixi}  \\
l_{E} &\equiv& -(c_{1}(2+p)n+2\alpha(1-2n)n)  \\
l_{S} &\equiv& -(c_{1}+c_{3})n(6n-p-10) .
\end{eqnarray}
where $k_{hub}\equiv\frac{1}{\eta_{today}}$.

The vector field affects our evolution equations for the matter and metric perturbations only through the above
two expressions. The first may be interpreted as proportional to an energy density contribution of the vector field (hence $l_{E}$) in the $G^{0}_{\phantom{0}0}$ Einstein equation, 
proportional
to both the co-moving wavenumber squared and the conformal time $\eta$ raised to the power $5+p-6n$.
The second contribution may be seen as an anisotropic stress due to the vector field (hence $l_{S}$) and has an identical 
$k$ and $\eta$ dependence as the energy density. Indeed, for general $c_{i}$ 
it is clear that the $l_{i}$ and thus both terms should be of similar order.

We will now characterize the possible evolution of the perturbations during the matter era 
for a wavenumber $k$ that is outside the horizon at the beginning of integration ($k\eta << 1$) 
and inside by the end ($k \eta >> 1$). Recall that in the absence of an vector field, 
and for suitable initial conditions, the potential and density contrast 
are constant outside the horizon, whereas inside the horizon the potential remains constant 
and the density contrast grows as $\eta^{2}$. In the presence of the vector field then 
the full system of evolution equations in the matter era is as follows:

\begin{eqnarray}
\label{full}
\delta'&=&-\theta+3\Phi' \\
\theta'&=&-{\cal H}\theta+k^{2}\Phi\\
\nonumber &&+l_{S}\xi(k) k^{2}\eta^{5+p-6n}\\
3{\cal H}\Phi'+k^{2}\Phi+3{\cal H}^{2}\Phi &=& -4\pi G\bar{\rho}a^{2}\delta  \\
\nonumber && +[3l_{S}(\frac{{\cal H}}{k})^{2}-\frac{1}{2}l_{E}]\xi(k) k^{2}\eta^{5+p-6n}  .
\end{eqnarray}

\subsection{$k\eta <<1$}

We assume adiabatic initial conditions for the fields $\delta,\Phi,\theta$.
In the super-horizon limit, one we may drop all terms which necessarily depend on positive powers of $k$ with respect to other terms.
Therefore the velocity divergence $\theta$ becomes decoupled from the evolution equations, which may then be combined to obtain a second order evolution equation for $\delta$:

\begin{eqnarray}
\delta''+\frac{6}{\eta}\delta'=6(5+p-6n)l_{S}\xi(k)\eta^{3+p-6n} 
\end{eqnarray}
with solution:

\begin{eqnarray}
\delta= C_{1}(k)+\frac{6l_{S}\xi(k)}{(10+p-6n)}\eta^{5+p-6n} 
\end{eqnarray}
where $C_{1}$ is a constant of integration and we have omitted the decaying mode.

Therefore even before horizon crossing, the anisotropic stress term due to the vector field can influence time evolution of the baryon density contrast to a degree dictated by the parameters of the theory and initial conditions on the vector field.

\subsection{$k \eta >>1$}

In the limit $k\eta >>1$ the $k^{2}\Phi$ term in the quantity $G^{0}_{\phantom{0}0}$ dominates the terms
with time derivatives, e.g. ${\cal H}\Phi'\sim \Phi /\eta^{2}$. The system of equations is given to a good approximation by:

\begin{eqnarray}
\delta'&=&-\theta+3\Phi' \\
\theta'&=&-\frac{2}{\eta}\theta+k^{2}\Phi-l_{S}\xi(k)(k\eta)^{2}\eta^{3+p-6n}  \\
k^{2}\Phi &\sim & -\frac{1}{2}l_{E}\xi(k)(k\eta)\eta^{3+p-6n}-\frac{6}{\eta^{2}}\delta
\end{eqnarray} 

Differentiating the density equation and substituting in for $\theta '$ using the velocity equation, 
we recover:

\begin{eqnarray}
\delta''&=&\frac{2}{\eta}\theta-k^{2}\Phi+l_{S}(k\eta)^{2}\eta^{3+p-6n}+3\Phi''\\
\nonumber &=& -\frac{2}{\eta}\delta'+\frac{6}{\eta}\Phi'-k^{2}\Phi+l_{S}(k\eta)^{2}\eta^{3+p-6n}+3\Phi'' 
\end{eqnarray}

As before we may neglect the contribution due to time derivatives of $\Phi$, then using the 
Einstein equation to substitute in for $k^{2}\Phi$ we finally recover:
 
\begin{eqnarray}
\delta''+\frac{2}{\eta}\delta'-\frac{6}{\eta^{2}}\delta=(\frac{1}{2}l_{E}+l_{S})\xi(k)(k\eta)^{2}\eta^{3+p-6n} 
\end{eqnarray}
with solution

\begin{eqnarray}
\label{dens}
\delta(k,\eta) &=& C_{2}(k)\eta^{2} \\
\nonumber      &&+\frac{(\frac{1}{2}l_{E}+l_{S})}{(5+p-6n)(10+p-6n)}\xi(k)(k\eta)^{2}\eta^{5+p-6n} 
\end{eqnarray}
where $C_{2}(k)$ is a constant of integration. 

\begin{figure}
\epsfig{file=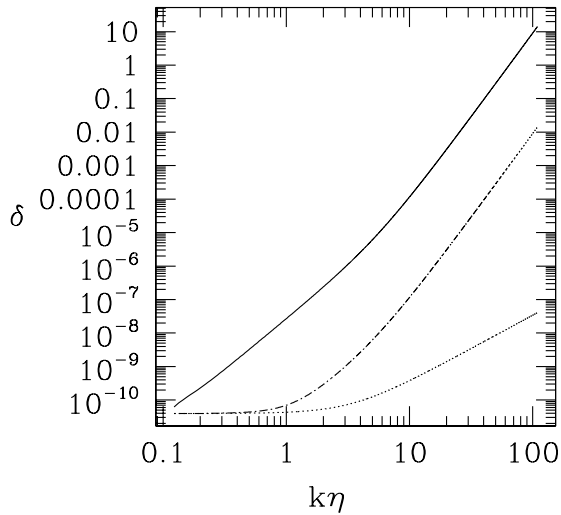,width=8.3cm,height=8.3cm}
\caption{Influence of the vector field on the growth of $\delta$. Dotted line is evolution of $\delta$ for a GR/baryon
universe. $\delta$ is constant until $k\eta~1$ and then grows as $\eta^{2}$. We now include the vector field. We choose $n=0.4$
and $p=0.4$. Dot-dash line shows the vector field initially subdominant to the matter perturbations. For $k\eta<<1$ it has negligible influence but at later times its energy density dominates and leads $\delta$ to grow as $\eta^{7+p-6n}=\eta^{5}$. The solid line shows a vector field dominant even at early times. Before horizon crossing the anisotropic stress dominates and leads to $\delta$ growing as $\eta^{5+p-6n}=\eta^{3}$. Eventually again the energy density dominates and $\delta$ grows as $\eta^{5}$ (parallel with dot-dash line)}
\label{fig3}
\end{figure}

We can see then that for sub-horizon modes, the influence of the vector field on the evolution of $\delta$ is a combination of the effect of the energy density and anisotropic stress contributions though both, in this limit, result in the same contributions to the scale dependence and time evolution of the density contrast. 
Two examples of the influence of the vector field on the evolution of a baryon overdensity mode are illustrated in figure (\ref{fig3}).

\section{The Power spectrum of matter 
perturbations}

Thus far we have explored the dynamics of the various instabilities
of both the vector field and baryon perturbations in generalized Einstein-Aether
theories. Under certain conditions, structure does grow. We can now
make a first attempt at calculating the power spectrum of baryon fluctuations
in a realistic universe. We will forego the complete analysis using
a full Einstein-Boltzmann code and focus on the lower order moments of
the matter field. Furthermore, we will not include Silk damping during
recombination at this stage. Like dark matter, the vector field is coupled to the other fields exclusively via source terms from the metric potentials and so as in the case of dark matter we expect the above approximation to be accurate
when looking to evaluate the linear baryonic matter power spectrum \cite{bma}.

The matter density can be split up into a radiation and baryonic component,
$\rho=\rho_B+\rho_\gamma$ with $\rho_B\propto a^{-3}$ and $\rho_\gamma \propto a^{-4}$.
We set equality, $\rho_B=\rho_{\gamma}$ at $a_{eq}= \Omega_{R}/\Omega_{m}$. We now split the
evolution of perturbation into two stages, before and after recombination at 
$a_*=10^{-3}$. Before recombination, the baryons and radiation are tightly
coupled through Thompson scattering and they can be described by a combined
fluid:
\begin{eqnarray}
\delta_{B\gamma}'&=&-\theta_{B\gamma}+3{\Phi'} \label{tc1} \\
\theta_{B\gamma}'&=& -\frac{\cal H}{1+R}\theta_{B\gamma}-\frac{R}{1+R}
\frac{k^2}{3}\delta_{B\gamma}+k^2\Psi \label{tc2}  ,
\end{eqnarray}
where $R=4\rho_\gamma/3\rho_{B}$.

After recombination, the two fluids decoupled and baryons will evolve as in section \ref{sect:vector}
while the radiation will obey:
\begin{eqnarray}
\delta_\gamma'&=&-\frac{4}{3}\theta_\gamma+4{\Phi'}  \label{radev1}\\
\theta_\gamma'&=&\frac{k^2}{4}\delta_{\gamma}+k^2\Psi \label{radev2}  .
\end{eqnarray}

At recombination, the matching conditions are: 

\begin{equation}
\delta_B=3\delta_\gamma/4=\delta_{B\gamma} \label{match1}
\end{equation}
and 

\begin{equation}
\theta_B=\theta_\gamma=\theta_{B\gamma} \label{match2}
\end{equation}. 

Therefore our approach is as follows: 

\begin{itemize}
\item For a given set of $\Omega_{b}$,$\Omega_{r}$,$H_{0}$,$c_{i}$,$n$, and $\gamma$ determine the background evolution
of the universe. We fix $h=0.73$ where $H_{0}=100h (km/s)Mpc^{-1}$ as well as $\Omega_{b}h^{2}=0.023$ and $\Omega_{r}h^{2}=2.47\cdot 10^{-5}$.
\item Choose initial conditions for the fields $\delta_{\gamma}$,$\theta_{\gamma}$,$\delta_{b}$,$\theta_{b}$,$\Phi$,$\Psi$,$V$, and $E$. We choose adiabatic initial conditions for fields other other than the vector field which, due to the smallness of ${\cal F}_{K}$, will be decoupled from the other fields at early times.
\item During tight coupling, numerically evolve the fields $\Phi$,$E$, and $V$ using the $G^{0}_{\phantom{0}0}$ Einstein equation and the vector field equation of motion while eliminating $\Psi$ via the constraint equation (\ref{GRTT}). Then evolve the matter fields using the tight coupling condition and equations (\ref{tc1}) and (\ref{tc2}).
\item At recombination use the matching conditions (\ref{match1}) and (\ref{match2}). Then continue to evolve the vector field and metric potentials as before but now evolve $\delta_{\gamma}$ and $\theta_{\gamma}$ using (\ref{radev1}) and (\ref{radev2}) while evolving $\delta_{b}$ and $\theta_{b}$ using (\ref{barev1}) and (\ref{barev2}). We cease integration at $z=0$.
\end{itemize}

The observable $\delta(k)$ today may be read from (\ref{dens}). It contains two functions which are determined in the numerical evolution: $C_{2}(k)$ and $\xi(k)$.
The form of the function $C_{2}(k)$ will follow from a given set of parameters and initial data.  Meanwhile, the function $\xi(k)$ is determined by the manner in which $V$ reaches a dominant growing mode solution.
We may isolate limiting, analytic cases for the modes of interest: The case in which the initial values of $E$ and $V/\eta$ are significantly smaller than the initial metric source terms (Case 1) and the case in which the initial values of $E$ and $ V/\eta$ are significantly greater than the initial metric source terms (Case 2).


\subsection{Case 1} \label{sect:Case1}

We first consider modes that enter the horizon after radiation contributions may be neglected. The scale dependence of these modes
may obtained to a good approximation by matching our super-horizon solution to our sub-horizon solution at horizon crossing. A given mode will enter the horizon when $k\eta_{ent}\sim 1$ where $\eta_{ent}$ is the conformal time at horizon crossing. Matching the solutions from the previous section we find the following solution for $C_{2}(k)$:

\begin{eqnarray}
C_{2}(k) &\simeq& \delta_{prim}(k)k^{2}+[\frac{6l_{S}}{(10+p-6n)} \\
\nonumber && -\frac{(\frac{1}{2}l_{E}+l_{S})}{(5+p-6n)(10+p-6n)}]\xi(k)k^{-3-p+6n}
\end{eqnarray}
where $\delta_{prim}(k)$ is the primordial amplitude of $\delta$.  If we assume adiabatic initial conditions and a scale invariant spectrum for $\Phi$ then $\delta_{prim}\propto k^{-3/2}$.  We must now find the scale dependence of $\xi(k)$. To do this it is necessary to examine how $V$ reached the monomial solution. To do this we solve (\ref{vecp}) for super-horizon modes during radiation domination, which reduces to:

\begin{equation}
V''+\frac{b_{1}}{\eta}V'+\frac{b_{2}}{\eta^{2}}V= \frac{l_{3}}{\eta}\Psi_{prim}
\end{equation}
where $l_{3}=-[\tilde{\alpha}(2n-1)+5-4n]$ .

We have assumed that at such early times the vector field does not affect 
superhorizon metric modes enough for them to deviate appreciably from their primordial values in determining the early evolution of $V$. The equation may be solved to yield:

\begin{equation}
\label{eq:supv}
V=D_{1}\eta^{p_{+}}+D_{2}\eta^{p_{-}}+\frac{l_{3}\Psi_{prim}}{b_{1}+b_{2}}\eta
\end{equation}

We now impose $V(k)/\eta_{0} << \Psi_{prim}$. 
This implies that the decaying mode coefficient $D_{2}$ should be very close to zero, and it shall be neglected from now on. Now differentiating the solution, the condition $V'(k)=E(k)<<\Psi_{prim}$ requires that

\begin{equation}
D_{1}=-\frac{l_{3}\Psi_{prim}(k)\eta_{0}^{1-p_{+}}}{p_{+}(b_{1}+b_{2})}
\end{equation}
and so

\begin{equation}
\label{eq:case1}
V=\Psi_{prim}(k)\eta \frac{l_{3}}{b_{1}+b_{2}}[1-\left(\frac{\eta}{\eta_{0}}\right)^{p_{+}-1}]
\end{equation}

Therefore, providing that $p_{+}-1>0$, the complementary function solution will tend to overcome the particular integral and the monomial solution will be reached. Then comparing with \ref{eq:simple} we find for these modes that:

\begin{equation}
\label{eq:vok}
V_{0}(k)=\Psi_{prim}(k)\frac{l_{3}}{b_{1}+b_{2}}\eta_{0}
\end{equation}

Thus, from equation (\ref{xixi}) we see that both $V_{0}(k)$ and $\xi(k)$ will have the same k-dependence
as $\Psi_{prim}(k)$. Putting everything together, we find that for modes entering the horizon after radiation domination that:

\begin{equation}
\delta^{2}(k,\eta)=f_{1}(\eta)k\quad\quad\quad (k<<k_{eq})  
\end{equation}
where the function $f_{1}(\eta)$ is determined by the solution to the field equations.

Therefore for modes that enter the horizon after radiation domination, the scale dependence of the power spectrum is expected to be 
identical to that of a pure baryonic or $\Lambda$CDM universe up to a multiplicative function of the conformal time. 

We now look at the behaviour of modes entering the horizon at a time when radiation contributions cannot be neglected. During radiation domination (again assuming a small stress-energy of the vector field at these early times) and before recombination the coupled photon-baryon fluid will undergo oscillations of constant co-moving amplitude. On an expanding background, the associated behaviour of the metric perturbation will be to undergo damped oscillations, these in turn sourcing the vector field equation. Once again we find that the vector field tends to the growing mode solution. As for the longer wavelength modes, the initial conditions dictate that the scale dependence of $V_{0}(k)$ (and thus the shape of the power spectrum if $V$ produces a dominant stress-energy perturbation) will be informed by the behaviour of the metric perturbations. From (\ref{vecp}) it can be seen that there will then be damped oscillatory terms in the vector field equation, there then, as in (\ref{eq:case1}), will be an integrated contribution to $V$ with a functional dependence on the numbers $c_{i}$ and $n$.  Indeed we find a range of parameters for which the damping of the metric potentials can become imprinted on $V_{0}(k)$. If $k^{2}V_{0}(k)$ is a decreasing function of k, then the influence of the vector field will diminish on small scales. 

In summary then, modes of the baryon density perturbation which enter the horizon while the metric perturbations are being damped may diminish in size with increasing $k$ at late times due to the effect of the damping of the metric perturbations on the evolution of the vector field, which, if yielding a realistic power spectrum, will at later times dominate the Poisson equation. Meanwhile, modes which enter the horizon after radiation may be neglected will grow as $k^{\frac{1}{2}}$ with increasing k. The demarkation between these regimes is the time of radiation-matter equality with an associated scale of the order $k_{eq}\sim \eta^{-1}_{eq}$ and we expect this to be observable in the resultant power spectrum. This is observable in figure (\ref{fig4})) which plots numerical solutions for the matter power spectrum.

\subsection{Case 2}

We now turn to the case where the initial conditions for $E$ and $V/\eta$ are considerably greater than metric perturbations. Recall that in the opposite case, we saw that the behaviour of the metric was instrumental in setting up the scale dependence of the growing vector mode. In this case, however, we will find that the eventual scale dependence of $V_{0}(k)$ is predominantly determined by our initial values of $E$ and $V/\eta$. Again we first consider modes which lie outside the horizon during radiation domination. The solution to $V$ will again be described by equation (\ref{eq:supv}). Again the decaying mode is neglected. Now though, we additionally may neglect the influence of the metric source. This imposes $E=V/\eta$ and:

\begin{equation}
\label{eq:supvv}
V=V(\eta_{0},k)\left(\frac{\eta}{\eta_{0}}\right)^{p_{+}}
\end{equation}

Clearly then, if we choose an initial condition $A^{i}(\eta_{0},k)\propto k^{-q}$ a dominant vector field will produce the following behaviour in $\delta(k)$ at late times for modes entering the horizon after radiation domination:

\begin{equation}
\label{eq:delsv}
\delta^{2}(k,\eta)=f_{2}(\eta)k^{4-2(q+1)}\quad\quad\quad (k<<k_{eq}) 
\end{equation}

where the $f_{2}(\eta)$ is determined by the solution to the field equations.

\begin{figure}
\epsfig{file=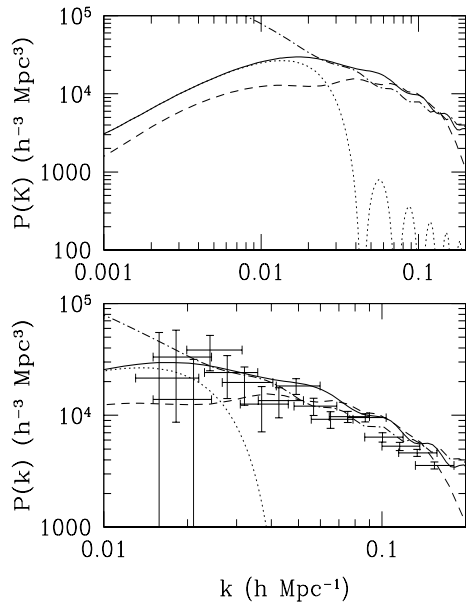,width=8.3cm,height=9.3cm}
\caption{Density contrast squared as a function of wavenumber at $z=0$. The dot-dash line is the $\Lambda$CDM prediction while the 
 dotted line is the prediction for a baryonic universe with $\Omega_{b}=0.27$. The solid and dashed lines represent baryonic universes
 $\Omega_{b}=0.04$ with a dominant vector field for cases 1 and 2 respectively. Data is from the Sloan Digital Sky Survey \cite{SDSS}}
\label{fig4}
\end{figure}

Crucially, the solution (\ref{eq:supvv}) in this case now even approximately to smaller wavelength modes entering the horizon during radiation domination: the metric source terms remain but are not significant in altering $V_{0}(k)$ and having suffered damping cannot appreciably influence the solution at least until the vector field begins to act a a dominant source of stress energy. Therefore,
the solution (\ref{eq:delsv}) is a good approximation across all modes of interest! In order to obtain a fit to the power spectrum on the most observationally constrained scales (0.01-0.1 Mpc$^{-1}$) we must take $q=-3/2$ i.e. to have the same scale dependence as $\Phi_{prim}$, and, by the fixed-norm constraint, $A^{0}$. Immediately then we can see that in this case the resulting power spectrum will diverge on large scales (see figure (\ref{fig4})).

\section{Comparison To TeVeS}

There exists another relativistic theory which reduces to MOND in the quasistatic, weak-field limit: the TeVeS theory proposed by Jacob Bekenstein (\cite{bek2}). As mentioned, it has been shown \cite{zfs1} that this theory can be re-written as a tensor-vector theory much like the model discussed in this paper, the significant difference being that the coefficients $c_{i}$ are now functions of $A^{\mu}A_{\mu}\equiv A^{2}$- the non-vanishing vacuum expectation value of $A^{2}$ appearing not via a constraint but dynamically via the $c_{i}(A^{2})$ functions diverging as $A^{2}\rightarrow0$. It has been found that it is also the growth of the vector field perturbation $V$ which is crucial in TeVeS for the formation of cosmic structure.

Denoting the cosmological background value of $A^{2}$ as $\bar{A}^{2}$ the field equation for $V$ in TeVeS in a regime
where $a\propto \eta^{m}$ is:

\begin{equation}
\label{eqn:teve}
V''+\frac{b_{1}}{\eta}V'+\frac{b_{2}}{\eta^{2}}V=S[\Phi,\Psi]
\end{equation} 
where now:

\begin{eqnarray}
b_{1} &=& m(3-\bar{A}^{4})\\ 
b_{2} &=& m(m-1)(2-\bar{A}^{4})\\
\nonumber &&+\bar{A}^{4}K_{B}^{-1}3m^{2}(1-\bar{A}^{4})
\end{eqnarray}
where $K_{B}$ is a constant of the theory. 

It was found in \cite{dodel} that during the matter era the quantity $\bar{A}^{2}$ is a function of time that will deviate very slightly from $-1$ towards $0$ as cosmic time increases. Again the metric source terms are initially subdominant during the matter era. If $K_{B}$ is comparitively large then its contribution to (\ref{eqn:teve}) is suppressed. In this limit $b_{1}\rightarrow 4$ and $b_{2}\rightarrow 2$ leading to decaying homogeneous solutions for $V$: $V \propto \eta^{-1}$ and $V \propto \eta^{-2}$ and the evolution eventually being dominated by the particular solution
$-\Psi\eta /3$.

Again we may choose the set of evolution equations where the contributions of the vector field to the evolution of other fields are characterized entirely by the energy density $-a^{2}\delta\tilde{T}^{0}_{\phantom{0}0}$  and anisotropic stress $k^{2}(\Psi-\Phi)$ (see \ref{comp}). These are given by:

\begin{eqnarray}
\label{comp}
-a^{2}\delta\tilde{T}^{0}_{\phantom{0}0} &=&-\frac{(1-\bar{A}^{4})}{\bar{A}^{4}}{\cal H}k^{2}V-\frac{K_{B}}{2}k^{2}E \\
k^{2}(\Psi-\Phi)&=&\frac{(1-\bar{A}^{4})}{\bar{A}^{4}}k^{2}(2{\cal H}V-E)
\end{eqnarray}

Therefore so the system of evolution equations will be identical to (\ref{full}) with contributions from the vector field
instead given by (\ref{comp}). Recall that in the absence of the vector field, the baryonic contribution to the Poisson
equation in the matter era is of the order $\delta/\eta^{2}$ (where we have used that fact that ${\cal H}=2/\eta$ during the matter era). The overdensity grows as $\eta^{2}$ and so the contribution is constant in time. In the event that the vector follows the particular solution proportional to $\eta$ we see that the contribution to the Poisson equation proportional to $E$ is also constant in time whereas the term proportional to ${\cal H}V$ grows very gradually due to the background behaviour of $A^{2}$. However, it was found that for modes entering the horizon prior to the end of radiation
domination that the value of $\Psi$ appearing in the particular solution was sufficiently damped compared to the primordial value that the influence of the vector field remained too small to produce realistic growth of structure.
Therefore, a more rapidly growing $V$ is required.

If $K_{B}$ is indeed though, the contribution to $b_{2}$ cannot be ignored. It acts to create a growing mode solution for $V$, with the exponent of $\eta$ itself a function of $a$, eventually becoming greater than 1 and dominating the particular solution. It was found that this more rapid growth does allow for the growth of structure, indeed providing enhanced
growth over the $\Lambda$CDM model.

However, it can be seen from the vector equation that the growing mode in the homogeneous solution only possible if $\bar{A}^{2}\neq -1$. Therefore, the fixed-norm constraint in and of itself is not necessary for structure formation in a tensor-vector theory. A second noteable difference occurs in the rate of growth of the stress-energy of the vector field. Recall that in the fixed-norm model, the time dependence of the background function $F_{K}$ may significantly alter the time evolution of the perturbed stress-energy whereas in TeVeS there is an analogous role played  by $(1-\bar{A}^{4})/\bar{A}^{4}$ \textit{and} a fixed number $K_{B}$, which itself must be small to allow a growing mode.

\section{$\Psi-\Phi$}

We can see from equation (\ref{GRTT}) that the difference in conformal Newtonian gauge potentials ($\Psi-\Phi$) may be sourced 
by the transverse,traceless component of the vector field's stress energy tensor. This component is proportional to
the combination $c_{1}+c_{3}$. Therefore we may expect the conformal Newtonian gauge potentials to only be equal if $c_{1}+c_{3}=0$.

Meanwhile, restricting ourselves to the parameter space where $V$ tends to the growing monomial solution requires
that the number $b_{3}$ in equation (\ref{eqn:sourcefree}) should be approximately zero. 

It can be seen from (\ref{bees}) though that $b_{3}=0$ is
generically satisfied if $c_{1}+c_{3}=0$ and $c_{2}=0$. However, with this combination the vector field has no influence
on the background evolution and thus no late time acceleration can occur. If we allow $c_{2}$ to be nonzero, 
we can have $b_{3}=0$ and $c_{1}+c_{3}=0$ only when the number $n$ is equal to $1/2$.
However, we have found that growing modes do not exist for this value in the matter era. Therefore, any growing monomial solution for $V$ comes from a set of $c_{i}$ such that $c_{1}+c_{3}\neq0$. Thus, from (\ref{GRTT}) there will generally be an accompanying source for $\Psi-\Phi$. 

This will lead the metric potentials $\Phi$ and $\Psi$ to differ in a scale dependent manner. This difference may be observable in physics at late times and at recombination. As discussed in (\cite{zhang},\cite{bert2},\cite{luestark}) at late times one may compare the deflection of light by large scale structure (an effect proportional to $\Phi+\Psi$) to the motions of galaxies in clusters (an effect proportional to $\Phi$), thus deducing $\Psi-\Phi$ on large scales. Additionally, under the assumption of adiabaticity we have that
the anisotropy in the CMB, $\Delta T({\hat n})/T$ in a given direction 
${\hat n}$ is given by
\begin{equation}
\label{eqn:anis}
\frac{\Delta T({\hat n})}{T}\simeq -\frac{1}{2}\Psi(\eta_*,d_*{\hat n}) -
\int_{\eta_*}^{\eta_0}d\eta'({\dot \Psi}+{\dot \Phi})[\eta',(\eta_0-\eta'){\hat n}]
\end{equation}
where $\eta_*$ is the conformal time of last scattering, $\eta_{0}$ is the conformal time today, and $d_*$ is the comoving radius of the
surface of last scattering. 


We can obtain a handle on the behaviour of the temperature anisotropy (\ref{eqn:anis}) on scales which remain outside the horizon today. We first consider the term proportional to the value of the potential $\Psi$ at recombination. Up to recombination, we assume that non-negligible contribution to the background energy density is due to the baryons and photons. In the absence of dark matter, radiation domination ends only \textit{after} recombination. In analyzing the evolution of the potentials up to recombination then we may neglect the baryonic contribution to the Einstein field equations.  Therefore $a\propto \eta$. Furthermore we will assume that we're at a time when the vector field has tended to the monomial solution for modes of interest and in doing so has not appreciably affected the other fields sourcing it (as one would expect from the early smallness of ${\cal F}_{K}$) and so the contribution to the vector field in the system of field equations will be given by equations (\ref{too}) and (\ref{sifif}) (with $m$ now equal to 1). We find this to be an accurate description of the evolution of superhorizon vector modes for the eventual production of realistic power spectra. As before, when considering superhorizon modes the velocities become decoupled from the field equations and terms explicitely proportional to positive powers of $k$ are discarded, yielding:

\begin{eqnarray}
\label{ack}
\delta_{\gamma}' &=& 4\Phi' \\
\frac{3}{\eta}\Phi'+\frac{3}{\eta^{2}}\Phi &=& -\frac{3}{2\eta^{2}}\delta_{\gamma}+3l_{S}\xi(k)\eta^{3+p-6n} 
\end{eqnarray}
where, again, the anisotropic stress due to the vector may be significant for these modes.
Differentiating the second equation we may use the first to replace the derivative of the density contrast with a derivative of the potential, thus obtaining a second-order equation for $\Phi$:

\begin{eqnarray}
\Phi''+\frac{4}{\eta}\Phi' &=& (5+p-6n)l_{S}\xi(k)\eta^{3+p-6n}
\end{eqnarray}
with solution

\begin{eqnarray}
\label{eqn:superho1}
\Phi(k,\eta_{*})= \Phi_{0}(k)+\frac{l_{S}}{8+p-6n}\xi(k)\eta_{*}^{5+p-6n}
\end{eqnarray}
where $\Phi_{0}$ is the primordial amplitude of $\Phi$ (it is identical to the integration constant as the vector field contribution vanishes as $\eta \rightarrow 0$). 

It immediately follows (using equation (\ref{sifif})) that:

\begin{equation}
\label{eqn:superho2}
\Psi(k,\eta_{*})=\Phi_{0}(k)+l_{S}\left(\frac{1}{8+p-6n}-1\right)\xi(k)\eta_{*}^{5+p-6n}
\end{equation}

Therefore the degree to which super-horizon modes deviate from the primordial $k$ dependence of the potentials at recombination depends upon the function $\xi(k)$. 
For Case 1 initial data, it can be seen from (\ref{eq:vok}) and (\ref{xixi}) that for modes outside the horizon during radiation domination, $\xi(k)$ will have the same scale dependence as the primordial value of $\Phi$. Assuming adiabadicity then we have $\xi(k)\propto\Phi_{prim}(k)\propto k^{-\frac{3}{2}}$. Thus in Case 1 the effect of anisotropic stress on superhorizon modes will be a time dependent, scale invariant rescaling of the primordial value. As can be seen in (\ref{eqn:anis}), a significant time dependence may produce a large scale CMB anisotropy today which is not approximately determined by primordial values (in contrast with the dark matter case). 
The effect is even more pronounced with Case 2 initial data where, by (\ref{eq:supvv}) and (\ref{xixi}), we see that the k-dependence of $\xi(k)$ is approximately given by the \textit{initial} k-dependence on $E$ and $V$. It was found that a reasonable fit was obtainable for the linear power spectrum for $V\sim k^{-5/2}$. However by (\ref{eqn:superho1}) and (\ref{eqn:superho2}) we see that this will lead to a correction to $k^{3}\Phi^{2}$,$k^{3}\Phi^{2}$,  on superhorizon scales which \textit{diverges} for small wavenumbers.

We now consider the integral contribution to (\ref{eqn:anis}), the so-called Integrated Sachs-Wolfe (ISW) \cite{wolfe}. It may also be shown that, allowing for the different form of dominant background energy density, similar expressions to (\ref{eqn:superho1}) and (\ref{eqn:superho2}) hold for superhorizon modes
during the matter era, the \textit{sum} of superhorizon potentials retaining a time dependence thus leading to a contribution to the integral. Recall that in $\Lambda$CDM cosmology, this integral becomes nonzero only after the onset of late time acceleration. Again the Case 1 field emulates the initial scale dependence of the metric while for the Case 2 field the effect of the anisotropic stress due to the vector field leads to a divergence on the largest scales. 

\begin{figure}
\epsfig{file=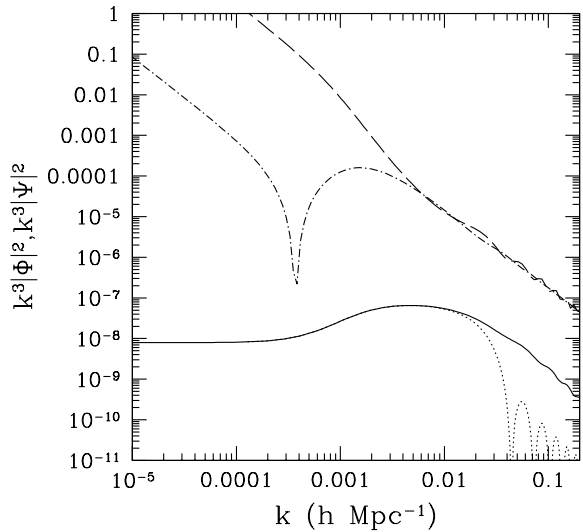,width=8.3cm,height=8.3cm}
\caption{Metric potentials as a function of wavenumber at $z=0$. Solid and dotted lines represent $k^{3}|\Phi|^{2}$ and $k^{3}|\Psi|^{2}$ each for a universe with ($\Omega_{DM}=0.23$,$\Omega_{b}=0.04$) (solid) and ($\Omega_{DM}=0.0$,$\Omega_{b}=0.27$) (dotted). Dashed and dot-dash lines represent $k^{3}|\Psi|^{2}$ and $k^{3}|\Phi|^{2}$ respectively for a universe with a Case 2 vector field.}
\label{fig5}
\end{figure}

The latter behaviour is illustrated in figure (\ref{fig5}) where potentials at $z=0$ are compared between the $\Lambda$CDM model and the Case 2 vector model depicted in figure (\ref{fig4}). The wavenumber dependence of the potentials in the  dark matter universe and Case 2 vector model agree for the range where the resulting baryon power spectrum is roughly similar. They are not of the same magnitude as in the Case 2 model the vector field has become the dominant energy density perturbation and so the matter perturbation is no longer simply related to the potential via the Poisson equation. For larger scales both potentials in the Case 2 model diverge from the dark matter case and from one another. As far as these initial conditions reflect the absence of turnover in the power spectrum, they also lead to $\Psi$, $\Phi$, and their sum diverging on large scales, these effects growing with conformal time due to the induced time dependence of the potentials. 

\section{Discussion and Conclusions} \label{sect:conc}

We have shown that the vector field in the Generalized Einstein Aether theory can play the same role as in TeVeS and the same role as dark matter in sustaining the growth of structure during
recombination. A growing mode for the vector field exists in the matter era for a wide variety of parameters and, through its associated energy density and anisotropic stress, can become a dominant source term in the Poisson equation, thus sustaining gravitational potentials which further act as sources for the evolution of the baryonic density contrast. The shape of the power spectrum is dictated predominantly by the $k$ dependence of the growing mode which may either be determined by dominant initial data for the vector field (Case 2) or the manner in which an initially very small vector field reaches the growing mode solution via coupling to evolving gravitational potentials (Case 1). As in the case of dark matter, a turnover in the power spectrum exists in the latter case due to the comparitive lack of damping of metric modes entering the horizon after radiation domination. 

Though the vector's stress energy tensor does contain terms proportional to the gravitational potentials (leading to an effective time dependent rescaling of Newton's constant in Poisson's equation) we find that this is a subdominant effect. At late times and on large scales Poisson's equation (\ref{eqn:Poisson})  takes the approximate form:

\begin{equation}
\label{eq:cosmol}
\nabla^{2}\Phi\sim 4\pi G\rho_{B}+\frac{l_{E}}{2}{\cal F}_{K}(\eta){\cal H}\nabla^{2}V
\end{equation}
where $\rho_{B}$ is the baryonic energy density.

The dominant component of the right hand side at late times is found to be the term associated with the field $V$. In so far as the evolution of $V$ is sourced by terms in the metric but not entirely fixed by it via constraints, its contribution to the Poisson equation is more akin to dark matter.

It is interesting to contrast this with the weak-field quasistatic limit of the theory discussed in \cite{zfs2}, wherein $A^{\mu}$'s stress energy tensor is determined \textit{entirely} by the metric and thus all observed mass discrepancies must be attributed to a modification of Poisson's equation which takes the form:

\begin{equation}
\label{eq:weakfi}
\nabla.(\mu(|\nabla\Phi|/M)\nabla\Phi)=4\pi G\rho_{B}
\end{equation}

Therefore in this regime mass discrepancies can be interpreted as arising from a change in the relationship between
the dominant component of the matter energy density (taken to be baryonic matter) and the gravitational field. 
In producing a realistic power spectrum, we find that the effective energy density of the growing mode $V$ is considerable at the scale where overdensities become nonlinear ($k\sim 2h Mpc^{-1}$). A similar situation exists in the TeVeS theory. This raises the issue, as yet unaddressed, of the degree to which $V$, instrumental in the growth of large scale structure, contributes to the mass of bound structure and, if so, whether it remains akin to a dark source in the Poisson equation (as in \ref{eq:cosmol}) rather than a modification of how the gravitational field is sourced by baryonic matter (as in \ref{eq:weakfi}). In other words, covariant realizations of the MOND program may well have unintentionally reintroduced dark matter,  albeit non-particulate dark matter, via the back door.
However, it is by no means obvious that the collective picture given by a universe where a vector field gives rise to mass discrepancies would be degenerate with a particle dark matter model i.e the theory may resemble different types of dark matter in different regimes but not a single type across all regimes. Indeed we have seen that the cosmological case involves significant anisotropic stresses at late times while the quasistatic case considered in  \cite{zfs2} does not. 

We were able to clarify the analogy to the cosmological perturbations of Bekenstein's TeVeS theory. While the vector field is also nonvanishing in the cosmological background in that case, it is seen that the growth of structure requires the vector field be of non-fixed norm. As in the model discussed here, a growing mode in the vector field eventually dominates the evolution of the baryonic matter, its scale dependence being determined by the influence of metric source terms in its equation of motion. However, in TeVeS the growing mode and vector field perturbed stress energy tensor will tend to have different time dependences and so make for a potentially differing evolution of perturbations over cosmic time. In both theories the quantity $\delta\tilde{T}^{0}_{\phantom{0}0}$ will generally retain a time dependence during the matter era, in contrast to the perturbed dark matter density in $\Lambda$CDM models.

We now consider the parameter space explored in this paper. We have found that a power spectrum with a turnover and 
giving a reasonable fit to data tend to follow when the vector field and its time derivative are initially small. Following this, the degree to which the power spectrum today can agrees with the data is dictated the values of the parameters $n$ and $\gamma$ (recall that ${\cal F}=\gamma(-{\cal K})^{n}$) and the $c_{i}$ (see section \ref{sect:Case1}). In the background we have seen (see figure \ref{fig1}) through that some values of $n$ and $\gamma$ in the modified Friedmann equation permit late time acceleration. To first order in perturbations we have seen that the influence of $n$ on the evolution of the matter and metric fields is via its combined influence on the exponent of the vector field growing mode and, along with $\gamma$, the time dependence of the quantity $F_{K}$ to zeroth and first order which themselves act as sources in the Poisson equation and difference between Newtonian gauge potentials. In particular we were able to find values of $n$ consistent with late time acceleration and a growing mode. We were unable to produce a realistic power spectrum in the absence of dark matter for the value $n=1$ (as hinted at by the lack of growing modes in the homogeneous solution to the vector equation for this value). 

The $c_{i}$, which determine the vector field's kinetic term, affect the degree to which the Friedmann equation is modified in the cosmological background. For instance, if $c_{1}+3c_{2}+c_{3}=0$ then there is no effect on the background evolution. To first order in perturbations, the $c_{i}$ determine the nature of growing mode. We have restricted ourselves to values of $b_{3}$ (see equation (\ref{bees})) which led to an approximate monomial growth of the vector field. Deviations from this behaviour will have an integrated impact on the evolution of the other fields again, for instance, acting as a dark source in the Poisson equation but it is to be expected that different $n,c_{i}$ will match observations of the matter power spectrum. 

It is worth noting that in this case as well as TeVeS a power-law growing mode emerges from a non-Maxwellian kinetic term (i.e. ${\cal K} \neq -F^{\mu\nu}F_{\mu\nu}$;$c_{1}+c_{3}\neq0$); it has been shown in \cite{zfs1,tartag,bf2} that such kinetic terms can arise from a change of variables at the level of the action. For instance in TeVeS the vector field kinetic term is Maxwellian when written as a bimetric theory but not when written as a single metric theory. Our choices for $F({\cal K})$ were dictated by a desire for simplicity. It is to be hoped that a deeper grounding of the ideas discussed herein would fix the expected form of ${\cal F}$. As all contributions of the vector field to the matter and metric evolution equations at 1st order in perturbations are proportional to the background value of ${\cal F}_{K}$, the time dependence of this quantity can be expected to have a significant impact. We have seen that this will tend to produce enhanced growth for viable forms of a monomial ${\cal F}(\cal K)$.  

Though we have chosen the sign of the term $\alpha=c_{1}+3c_{2}+c_{3}$ to be consistent with subluminal propagation of gravitational waves in the limit far from matter, in generic backgrounds the speed of gravitational waves will depend on $F({\cal K})$ and it is not obvious that the theory would pass tests such as those discussed in \cite{moore}.
Recent work \cite{li,garfinkle} on the consequences of the theory with $F \propto {\cal K}$ have chosen the opposite sign of $\alpha$ where gravity propagates superluminally with respect to the preferred frame, thus avoiding any constraints from the above tests \cite{PPN} ; such a choice would change the requirements for a growing vector mode for more complicated $F({\cal K})$ as well as leading to the background cosmology and weak field limit being described by the same sign ${\cal K}$, rendering them no longer independent regimes.

Finally, the smallness of the mass scale appearing in the vector action $M\sim cH_{0}<<M_{Pl}$ is purely phenomenological. Though the value lends itself to modifications of gravity associated with low gravitational fields in the weak field limit and at late times in the cosmological background (i.e. as $H\rightarrow M$), it is fixed as a number in the action entirely by the former and lacks a deeper theoretical justification. A model wherein $M$ itself is a function of cosmic time (through, say, a background time dependence on $A^{2}$) may have considerable effects on the background cosmology as well as time evolution of the perturbed vector field’s stress energy tensor. 

As the size of perturbations is constrained on large scales by the baryon power spectrum at late times and the CMB anisotropy at recombination (allowing for the ISW effect), a full simulataneous modelling of each in the presence of the vector field should allow further discrimination between models and the degree to which they can give an account of mass discrepancies compatible with data in precision cosmology.

In summary we have considered a set of models of a universe with a timelike vector field with a noncanonical kinetic term. We have found that the model on one hand exhibits behaviour typical of `modified gravity' theories in high symmetry (for instance modifying the Friedmann equation in the FRW background and the Poisson equation in the static spherically symmetric case \cite{zfs2}). Additionally we have found that the extra degrees of freedom introduced through covariant realization of the former behaviour behave in a manner more akin to dark matter in perturbations around the FRW background. In particular, we find that at late times the stress energy tensor components of the field $V$ can become the dominant component in the cosmological Poisson equation. It is not clear, for instance, that this dominant component should be found to align with displacement of the subdominant baryonic matter in large scale structure. Even in so far as the field $V$ may resemble dark matter in the linear regime we have found a number of aspects in which its account differs from that of cold dark matter. We hope that this may help point the way towards a general approach to observationally distinguishing between the concordance model of cosmology and modified gravity.

{\it Acknowledgments}:
We thank Ted Jacobson, Constantinos Skordis, and David Mota for discussions.
TGZ is supported by a PPARC studentship. GDS was supported by Guggenheim and Beecroft Fellowships, and by
a grant from the US Department of Energy to the particle-astrophysics
theory group at CWRU.  GDS and PF thank the Galileo Galilei Institute
of Florence for their hospitality.  GDS thanks Oxford Astrophysics
for its hospitality, and Mapelsoft for the use of Maple software.

\end{document}